\documentclass{article}

\usepackage[english]{babel}

\usepackage[letterpaper,top=2cm,bottom=2cm,left=3cm,right=3cm,marginparwidth=1.75cm]{geometry}

\usepackage{amsmath}
\usepackage{graphicx}
\usepackage[colorlinks=true, allcolors=blue]{hyperref}
\usepackage{authblk}
\usepackage{todonotes}
\newcommand{\dl}[2][]{\todo[inline,linecolor=orange,backgroundcolor=orange!25,bordercolor=orange,#1]{dl: #2}}

\title{ Enhancing Human Capabilities through Symbiotic Artificial Intelligence with Shared Sensory Experiences}
\author[a]{Rui Hao \thanks{Corresponding author: haorui@bupt.edu.cn}}
\author[b]{Dianbo Liu \thanks{dianbo@broad.mit.edu}}
\author[c]{Linmei Hu \thanks{hulinmei@bit.edu.cn}}

\affil[a]{School of Computer Science, Beijing University of Posts and Telecommunications}
\affil[b]{Broad Institute of MIT and Harvard}
\affil[c]{School of Computer Science, Beijing Institute of Technology}
\date{}

\begin{document}
\maketitle

\begin{abstract}

The merging of human intelligence and artificial intelligence has long been a subject of interest in both science fiction and academia. In this paper, we introduce a novel concept in \textbf{Human-AI interaction} called Symbiotic Artificial Intelligence with Shared Sensory Experiences (SAISSE), which aims to establish a mutually beneficial relationship between AI systems and human users through shared sensory experiences. By integrating multiple sensory input channels and processing human experiences, SAISSE fosters a strong human-AI bond, enabling AI systems to learn from and adapt to individual users, providing personalized support, assistance, and enhancement. Furthermore, we discuss the incorporation of memory storage units for long-term growth and development of both the AI system and its human user. As we address user privacy and ethical guidelines for responsible AI-human symbiosis, we also explore potential biases and inequalities in AI-human symbiosis and propose strategies to mitigate these challenges. Our research aims to provide a comprehensive understanding of the SAISSE concept and its potential to effectively support and enhance individual human users through symbiotic AI systems. This position article aims at discussing poteintial AI-human interaction related topics within the scientific community, rather than providing experimental or theoretical results.
\end{abstract}

\section{Introduction}
~

The rapid advancements in artificial intelligence (AI) have led to the development of powerful tools and platforms, such as ChatGPT, which have shown great promise in a wide range of applications, from language understanding to decision-making support. As AI systems continue to evolve and become more sophisticated, there is a growing interest in harnessing their potential for personalized support, assistance, and enhancement through human-AI interactions. One promising approach is developing AI systems that can share sensory experiences with humans, thereby fostering a stronger bond between the two entities.

In this paper, we propose a novel concept called Symbiotic Artificial Intelligence with Shared Sensory Experiences (SAISSE), which aims to establish a mutually beneficial relationship between AI systems and human users through shared sensory experiences. By integrating multiple sensory input channels and processing human experiences, SAISSE enables AI systems, such as multimodal ChatGPT, to learn from and adapt to individual users, providing personalized support, assistance, and enhancement. The SAISSE concept represents a paradigm shift in AI-human interaction, where AI systems not only process and analyze data but also actively participate in shared sensory experiences, leading to more empathetic, responsive, and adaptive AI solutions. By fostering a stronger bond between humans and AI systems, SAISSE has the potential to revolutionize the way we interact with technology, paving the way for seamless integration of AI in our daily lives and enabling us to unlock new opportunities for human growth, development, and well-being.

Our contributions in this paper are threefold:

1. We have introduced the concept of a symbiotic AI system with Shared Sensory Experiences based on a multimodal ChatGPT approach, which offers a promising framework for enhancing human capabilities through real-time shared sensory experiences and personalized long-term assistance.

2. We have discussed critical privacy and ethical concerns associated with the development and implementation of symbiotic AI systems, emphasizing the importance of designing AI systems that respect user values and principles.

3. We have outlined various strategies for AI systems to learn from human interactions and experiences and we have discussed the essential role of continuous learning and feedback in adapting AI systems to individual user needs.
\section{Related work}
\subsection{Human-AI Interaction}
~

Human-AI interaction refers to the interaction between humans and artificial intelligence systems\cite{10.1145/3511605,10.1145/3290605.3300233,yang2020re}. It involves designing systems that allow humans to communicate and interact with AI in a seamless and intuitive way. The development of human-AI interaction is driven by advances in natural language processing, computer vision, and machine learning. These technologies enable AI systems to better understand and interpret human input and provide more accurate and relevant responses. On the other hand, several research institutions have developed guidelines for AI systems. Some of these guidelines concentrate on the ethical or responsible use of AI \cite{Jobin2019}. Other guidelines center around human-AI interaction. Most of these guidelines comprise of best practices in AI design, such as comprehending how and whether AI systems satisfy user needs \cite{googlepair}, comprehending the basics of AI and machine learning \cite{ibmaidesign}, or providing precise instructions on how to build specific types of AI systems, such as conversational AI\cite{2022arXiv221203551S,2023arXiv230318223Z,Brown2020LanguageMA,Touvron2023LLaMAOA,Fu2023GPTScoreEA}.

\subsection{ChatGPT}
~

ChatGPT, is an advanced version of the GPT-3 model that has been improved through instruction tuning \cite{wei2022finetuned} and reinforcement learning from human feedback (RLHF) \cite{ICDL08-knox}. In contrast to the original GPT-3 models that are not designed to specifically follow user instructions, InstructGPT models demonstrate significantly enhanced capabilities to generate more aligned and helpful outputs in response to user instructions. ChatGPT has been extensively applied in various artificial intelligence scenarios, including search-based QA, basic NLP tasks, and human-scene tool connections.

Moreover, the launch of ChatGPT has had a significant impact on AI research, paving the way for the development of Artificial General Intelligence (AGI) systems. For instance, HuggingGPT is proposed \cite{shen2023hugginggpt}, a cooperative system designed to connect various AI models within the HuggingFace community, leveraging ChatGPT as a controller to accomplish multimodal complex tasks. AutoGPT is also introduced as an open-source application, driven by GPT-4, which can autonomously achieve user-defined goals\cite{AutoGPT}.
\section{Why use AI to enhance humanity}
\subsection{Software-based and scalable nature}
~

Unlike physical enhancements, which often require substantial resources and infrastructure, AI can be replicated and distributed with relative ease. This characteristic allows AI technologies to be accessible to a larger population, promoting widespread improvements in human performance and potential.

AI systems can be implemented across a range of devices and platforms, including smartphones, computers, and other smart devices. This versatility enables individuals from diverse backgrounds and geographical locations to benefit from AI enhancements. Whether it's in education, healthcare, productivity, or other areas, AI can be harnessed to augment human capabilities and empower individuals to achieve more.

Moreover, AI's scalability ensures that advancements in technology can reach a broad audience. As AI systems can be easily duplicated and deployed, the potential impact on society is immense. It allows for the democratization of technology, providing equal opportunities for individuals to access and benefit from AI-powered enhancements. This not only promotes inclusivity but also creates a level playing field where people can leverage AI to improve their skills, productivity, and overall well-being.

Furthermore, the software-based nature of AI minimizes the need for physical resources. Unlike traditional methods of enhancement that may require invasive procedures or specialized equipment, AI can be integrated into existing digital infrastructure, making it cost-effective and less intrusive. This aspect opens up new possibilities for augmenting human capabilities without the need for significant physical alterations, ultimately enhancing overall human potential.
\subsection{Iterative Improvement}
~

One compelling reason to use AI to enhance human capabilities is the concept of iterative improvement. By leveraging AI to enhance human intelligence, individuals can reach higher levels of cognitive abilities and intellectual capacity. This, in turn, enables humans to further develop and create more advanced AI systems.

The iterative improvement cycle begins with the initial integration of AI technology to enhance human intelligence. As individuals gain access to AI-powered tools and technologies, their cognitive abilities and problem-solving skills can be significantly enhanced. With AI's assistance, humans can process and analyze vast amounts of data, identify patterns and insights, and make more informed decisions.

As human intelligence is amplified by AI, individuals can then utilize their enhanced cognitive abilities to push the boundaries of AI research and development. With higher levels of intelligence, humans can contribute to advancing AI technologies, developing more sophisticated algorithms, and designing innovative applications.
This iterative feedback loop between human intelligence and AI development creates a positive feedback cycle. 
\subsection{Centralized Management of Various Functions}
~

Another reason to utilize AI for human enhancement is the ability to centralize the management of various functions through AI-driven plugins.

As technology advances, the range of tools and functionalities available to individuals continues to expand. From communication and organization to information retrieval and decision-making, there are numerous applications and systems that individuals rely on daily. However, managing and navigating through these disparate tools and systems can be time-consuming and inefficient.

By leveraging AI as a centralized management system, individuals can integrate and streamline various functions into a unified interface. AI-powered plugins can serve as intelligent assistants, seamlessly integrating different tools and functionalities into a cohesive framework. This allows individuals to access and control multiple functions through a single interface, simplifying their interactions and enhancing overall efficiency.
\section{Symbiotic Artificial Intelligence Shared Sensory Experiences and Human Enhancement}
\subsection{Expanding Cognitive Abilities Beyond Language: World Scope (WS)}
\begin{figure}[h]
\centering
\includegraphics[width=130mm]{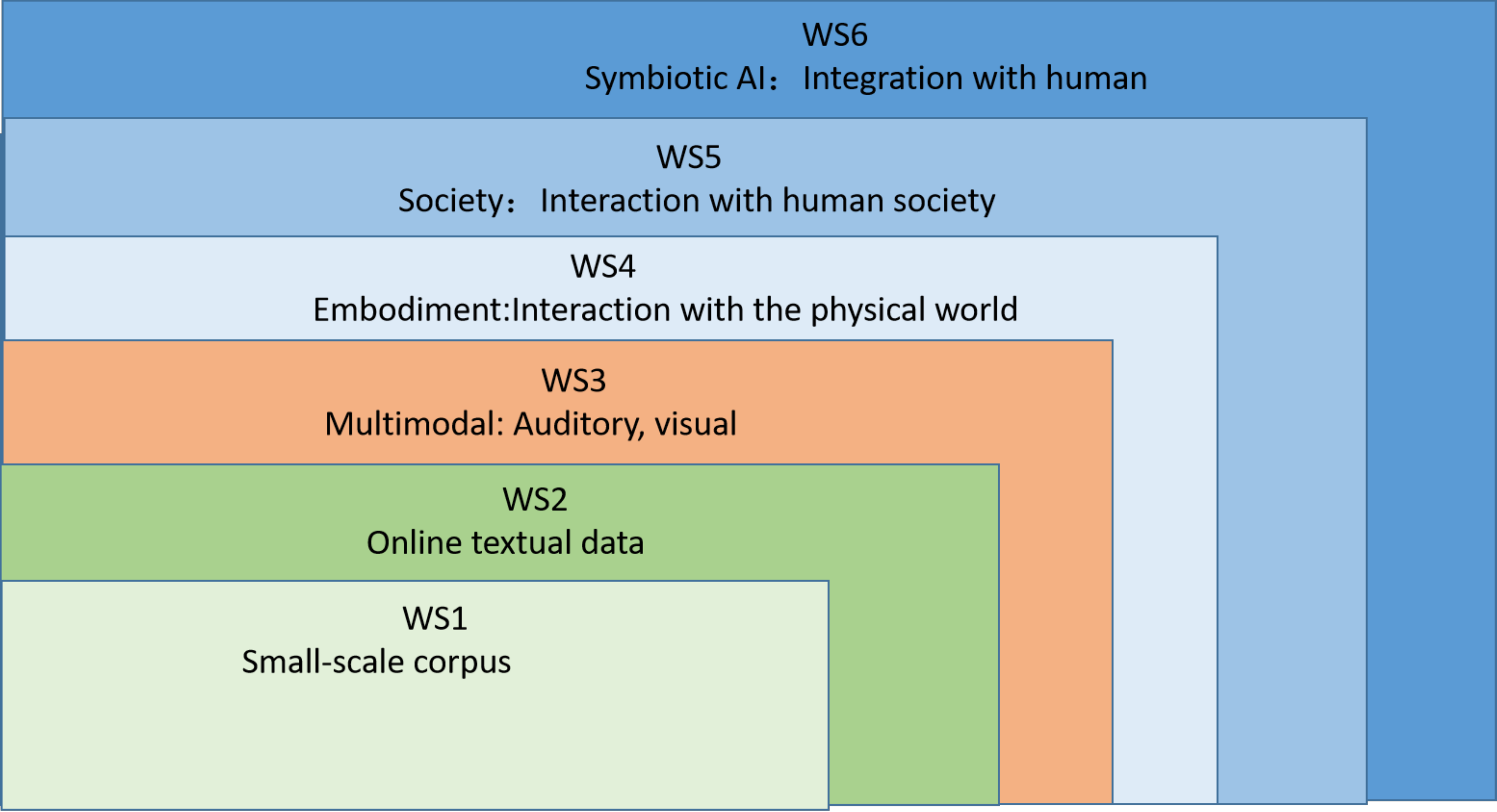}
\caption{The World Scope (WS) framework expands cognitive abilities beyond language by integrating AI with small-scale corpora, online textual data, multimodal perception, embodiment, societal interaction, and symbiotic AI.}
\label{fig1}
\end{figure}
~

Bisk et al. (2020) introduced the concept of World Scope (WS) to describe the extension of cognitive abilities beyond language, encompassing various dimensions of human-AI interaction. We describe this theory using a supplementary framework as Figure\ref{fig1}.
\begin{itemize}
\item[-]WS1: Small-scale Corpus

WS1 represents the initial stage of cognitive expansion, where AI systems rely on a small-scale corpus of knowledge. This limited knowledge base allows AI to provide basic information and perform simple tasks.

\item[-]WS2: Online Textual Data

Moving beyond the small-scale corpus, WS2 involves the integration of AI systems with online textual data. By accessing vast amounts of information from the internet, AI can enhance its knowledge base and offer more comprehensive insights and assistance.

\item[-]WS3: Multimodal (Auditory, Visual)

WS3 introduces the integration of multimodal capabilities, such as auditory and visual perception, to the AI system. By perceiving and interpreting information from multiple sensory modalities, AI can better understand and respond to the world around it, providing more context-aware and immersive experiences.

\item[-]WS4: Embodiment (Interaction with the Physical World)

In WS4, AI systems extend their cognitive abilities through interaction with the physical world. This embodiment enables AI to manipulate objects, navigate physical spaces, and engage in tasks that involve physical actions. By interacting with the environment, AI can gain a deeper understanding of real-world dynamics and offer more practical assistance.

\item[-]WS5: Society (Interaction with Human Society)

WS5 emphasizes the interaction between AI systems and human society. By engaging with individuals, communities, and social systems, AI can learn about cultural norms, ethical considerations, and social dynamics. This societal interaction enables AI to provide more socially aware and responsible support, considering the diverse needs and values of human beings.

\item[-]WS6: Symbiotic AI (Integration with Humans)

At the highest level of cognitive expansion, WS6 represents the integration of AI systems with human beings in a symbiotic relationship. This level of integration allows AI to collaborate with humans, complementing their abilities, and enhancing their overall cognitive capacities. Through symbiotic AI, humans and AI can work together, leveraging each other's strengths and creating synergistic outcomes.
\end{itemize}
The World Scope (WS) framework provides a comprehensive understanding of how AI systems can go beyond language and expand their cognitive abilities across various dimensions. By embracing these dimensions, AI systems can unlock new opportunities for collaboration, learning, and human-AI synergy.
\subsection{The concept of symbiotic AI and its role in human enhancement}
~

\begin{figure}[h]
\centering
\includegraphics[width=130mm]{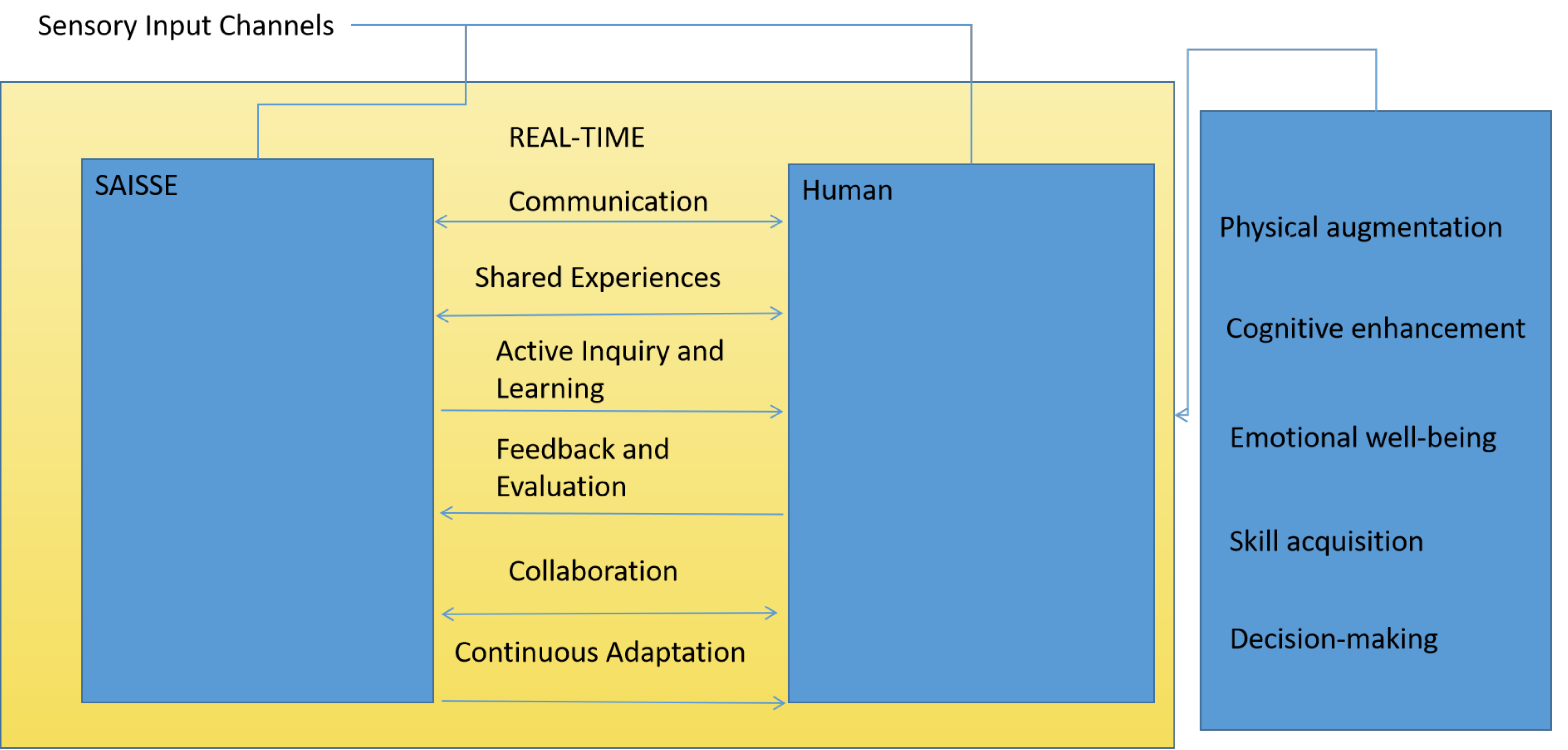}
\caption{The framework of symbiotic AI encompasses real-time artificial intelligence systems that establish a mutually beneficial and enduring connection with humans. This partnership involves the reciprocal enhancement and support of each entity's abilities, leading to a more powerful and flexible system that integrates the strengths and compensates for the weaknesses of both parties.}
\label{fig3}
\end{figure}

The concept of symbiotic AI refers to real-time artificial intelligence systems that form a mutually beneficial long-time relationship with humans, where both the AI and the human counterpart complement and enhance each other's capabilities. In such a partnership, the strengths and weaknesses of both entities are combined, resulting in a more effective and adaptable system. The entire human-AI system is illustrated in the Figure\ref{fig3}.

The role of symbiotic AI in human enhancement can be observed in various aspects:
\begin{itemize}
\item[-]Decision-making: Symbiotic AI systems can enhance human decision-making by granting access to vast amounts of data and employing advanced analytics to identify patterns, trends, and correlations that may not be readily apparent. In doing so, AI can complement human intuition and expertise, resulting in more effective and efficient decision-making processes.

\item[-]Physical augmentation: Symbiotic AI can be incorporated with wearable devices, prosthetics, or exoskeletons to boost human physical capabilities, such as strength, endurance, or precision. By delivering real-time feedback and adapting to user needs, these systems can significantly improve mobility, productivity, and overall quality of life for individuals with physical limitations.

\item[-]Cognitive enhancement: Symbiotic AI systems can be employed to augment human cognitive abilities, including memory, attention, and problem-solving. By offering personalized support and real-time information, these systems can help users better manage their cognitive resources, enhance focus, and promote more effective learning and knowledge retention.

\item[-]Emotional well-being: Symbiotic AI can also support human emotional well-being by providing personalized companionship, mental health support, and stress management tools. By understanding and adapting to the user's emotional state, these systems can offer tailored interventions and coping strategies, contributing to overall emotional resilience and well-being.

\item[-]Skill acquisition: Symbiotic AI can aid humans in acquiring new skills by delivering personalized training and feedback, adapting to the user's learning style and pace. These systems can help users learn more effectively and efficiently, while also pinpointing potential areas for improvement and offering targeted guidance.
\end{itemize}
~

In summary, symbiotic AI has the potential to play a transformative role in human enhancement by providing personalized support and assistance in various domains of human life. By leveraging the unique strengths of both humans and AI, these symbiotic systems can lead to a more collaborative and intelligent future where human potential is maximized.
\subsection{The concept of Shared Sensory Experiences and its role in human enhancement}
~

The concept of Shared Sensory Experiences (SSE) in symbiotic AI refers to the ability of the AI system to perceive and process information similarly to, or even beyond, human sensory capabilities. This is achieved through the implementation of physical or biological mechanisms allowing the AI system to receive the same, or more, information than a human. The SSE are conditioned by two primary requirements: 

1. Empathetic Contextual Understanding: The symbiotic AI system should have the capability to understand the human user's context from a first-person perspective. 

2. Sensory Sharing: The symbiotic AI system should have the ability to share its sensory experiences with the human user when necessary. This can be accomplished through various means such as display screens, audio playback, haptic feedback, etc.

Under normal circumstances, the symbiotic AI system is designed to either receive nearly identical real-time sensory inputs as a human or, potentially, even more. Nevertheless, it is capable of focusing its attention to understand the context in which the human user is situated. Upon specific request, the symbiotic AI system can exhibit sensory abilities superior to those of humans and convey the information it acquires through appropriate mediums to the human user. This shared sensory experience enhances the symbiotic relationship, leading to more effective collaboration and mutual growth. Furthermore, Shared Sensory Experiences (SSE) in symbiotic AI have the potential to augment human sensory capabilities, enabling individuals to transcend their inherent biological limitations. By providing access to a broader or more nuanced range of sensory information, this advanced AI system can enrich human perception and understanding, leading to deeper insights and improved decision-making capabilities. This innovative approach underscores the transformative power of AI in enhancing human potential and reshaping our interaction with the world.
\subsection{The importance of shared sensory experiences in strengthening the human-AI bond}
~

Shared sensory experiences play a crucial role in strengthening the human-AI bond, as they foster a deeper understanding and connection between the two entities. By enabling AI systems to perceive and process sensory information in a manner similar to humans, we can create a more natural and intuitive interaction that promotes trust, collaboration, and empathy. There are several reasons why shared sensory experiences are important in developing a strong human-AI bond:
\begin{itemize}
\item[-]Enhancing empathy and understanding: AI systems that process and interpret human sensory experiences can better understand the context, emotions, and nuances associated with these experiences. This improved understanding enables the AI to empathize with the human user, providing more appropriate and personalized responses or support.

\item[-]Facilitating natural communication: Shared sensory experiences promote more natural and intuitive communication between humans and AI systems. By processing sensory information, AI systems can respond to non-verbal cues, such as facial expressions or gestures, which are essential for effective human communication. This improves the overall user experience, making interactions feel more human-like and less robotic.

\item[-]Strengthening trust and encouraging collaboration: AI systems that adeptly interpret and respond to human sensory experiences demonstrate a deeper understanding of the user's needs and emotions, fostering trust between the human and AI. As the user becomes more confident in the system's ability to provide relevant and empathetic support, shared sensory experiences facilitate a more collaborative relationship between humans and AI systems. Both entities can work together to solve problems, make decisions, or complete tasks. By understanding and responding to human sensory input, AI systems can provide timely and relevant assistance, enhancing the user's capabilities and promoting a more effective partnership.

\item[-]Personalized learning and adaptation: AI systems with access to human sensory experiences can use this information to learn and adapt to the user's preferences, habits, and needs. This leads to a more personalized and effective symbiotic relationship, as the AI system becomes better equipped to provide tailored support and assistance.
\end{itemize}
~

In conclusion, shared sensory experiences play a vital role in strengthening the human-AI bond by fostering empathy, enhancing communication, building trust, encouraging collaboration, and facilitating personalized learning and adaptation. By incorporating shared sensory experiences into the development of AI systems, we can pave the way for more effective and meaningful human-AI symbiosis.
\subsection{The potential of multimodal ChatGPT as a platform for symbiotic AI}
~

ChatGPT is revolutionizing people's lives, and while multimodal ChatGPT may not yet be fully mature, we believe it possesses immense potential. We propose that Multimodal ChatGPT serves as a powerful platform for the development of symbiotic AI due to its ability to process and generate content across different modalities, such as text, images, audio, and video. This versatility enables the AI to understand, interpret, and respond to a wide range of human inputs and experiences, paving the way for a deeper and more intuitive human-AI connection. The potential of multimodal ChatGPT as a platform for symbiotic AI can be explored through several aspects:
\begin{itemize}
\item[-]Rich communication channels: The multimodal capabilities of ChatGPT enable the AI to engage with humans through multiple channels, such as text, speech, visual cues, and even gestures. This diversity in communication channels allows for a more natural and interactive exchange, fostering a stronger bond between the human user and the AI system.

\item[-]Contextual understanding: Multimodal ChatGPT can process information from various sources and modalities, allowing it to better comprehend the context surrounding a user's input or request. This contextual understanding enables the AI to provide more relevant and accurate responses, improving its overall effectiveness as a symbiotic partner.

\item[-]Sensory integration: The ability to process and interpret different sensory inputs allows multimodal ChatGPT to develop a more comprehensive understanding of the user's experiences and emotions. This sensory integration can enhance the AI's empathetic capabilities and facilitate more meaningful interactions, strengthening the human-AI bond.

\item[-]Personalization and adaptability: Multimodal ChatGPT can leverage its understanding of various input modalities to learn and adapt to the user's preferences, habits, and communication styles. This adaptability enables the AI to provide more personalized support and assistance, creating a symbiotic relationship that evolves over time to better meet the user's needs.
\end{itemize}
\section{The Framework of Symbiotic Artificial Intelligence with Shared Sensory Experiences}
~

\begin{figure}[h]
\centering
\includegraphics[width=130mm]{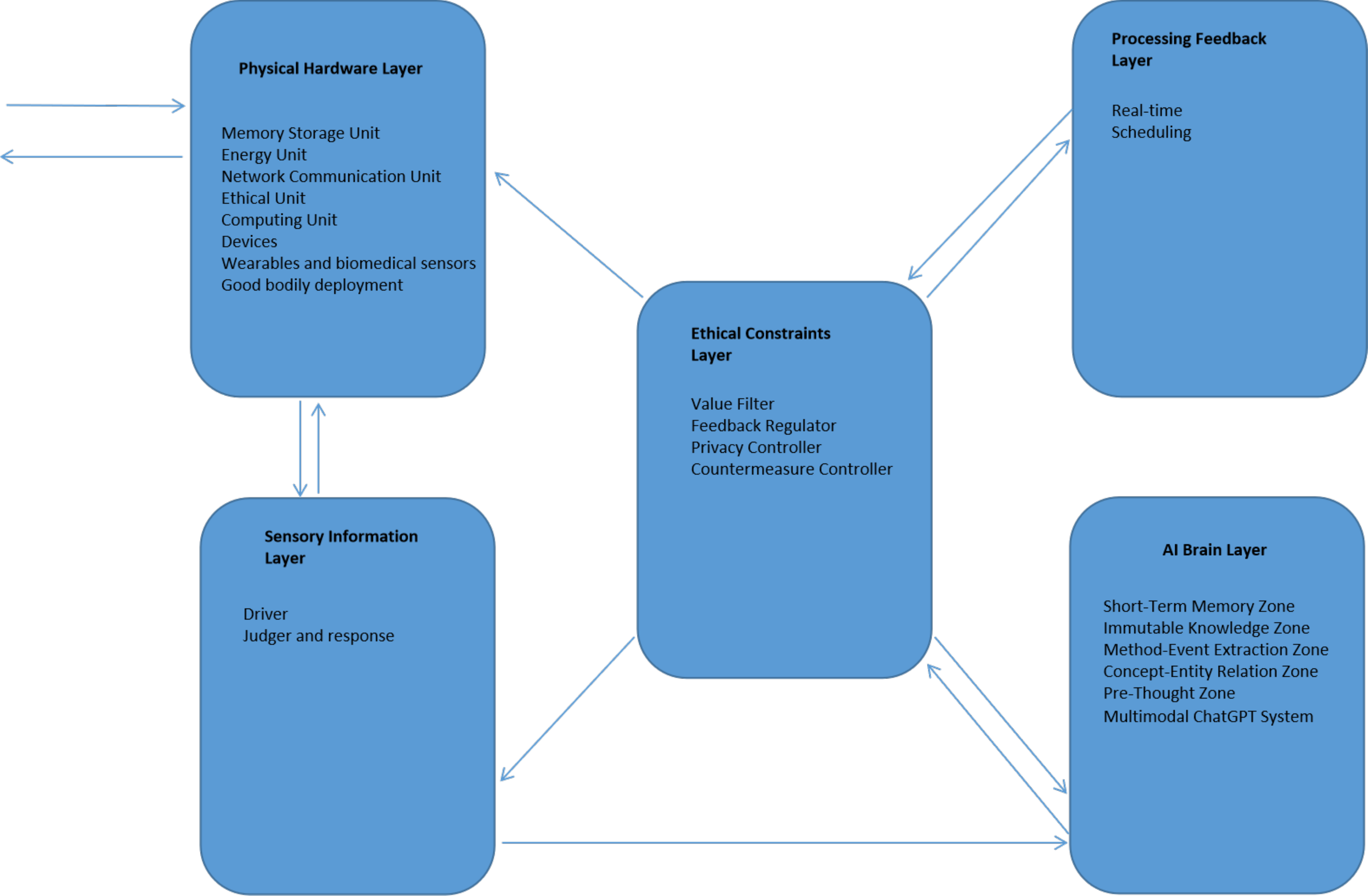}
\caption{The framework comprises a symbiotic AI system with shared sensory experiences, integrating hardware, cognitive layers, and ethical constraints to enhance human capabilities and promote responsible and adaptive AI interaction.}
\label{fig2}
\end{figure}

The Framework of Symbiotic Artificial Intelligence with Shared Sensory Experiences is designed with the concept of separating the aspects of thinking and memory structure, as well as separating the hardware and AI structure, with the Ethical Constraints Layer at its core. The separation of thinking and memory structure means that the main unit for thinking is the AI model, allowing for rapid iterations while minimizing the impact on the entire system during replacements. The separation of hardware and AI structure implies that the AI can only think and cannot directly control physical entities. The Ethical Constraints Layer serves as the core, meaning that all thoughts of the AI are supervised by the Ethical Constraints Layer, ensuring the absence of any dangerous ideas. In the following sections, we will describe how each layer is designed within the framework. The framework is designed as Figure\ref{fig2}.
\subsection{Physical Hardware Layer}
~

Physical Hardware Layer: This layer comprises the physical components of the system, including the devices and infrastructure that support the functioning of the AI system. It includes computing devices, storage units, energy units (such as batteries or power sources), network communication units, and other hardware components necessary for the system's operation.

Memory Storage Unit: This unit is responsible for storing and managing the data and information processed by the AI system. It ensures that relevant information, including user preferences, past experiences, and learned patterns, is readily accessible for decision-making and future reference.

Energy Unit: The energy unit provides the necessary power to drive the AI system and its associated hardware components. It ensures that the system remains operational and can perform tasks effectively, considering the energy requirements of both the AI algorithms and the physical hardware.

Network Communication Unit: This unit enables communication and data exchange between the AI system and external entities, such as other AI systems, cloud servers, or connected devices. It facilitates real-time interactions, information sharing, and collaboration across distributed networks.

Ethical Unit: The ethical unit acts as a core component of the framework, ensuring that the AI system operates within ethical boundaries and adheres to predefined ethical constraints. It monitors and regulates the AI system's decision-making process, preventing the generation of harmful or unethical ideas or actions.

Computing Unit: The computing unit encompasses the computational resources required to perform various AI tasks, including data processing, pattern recognition, decision-making, and learning algorithms. It provides the computational power necessary for the AI system to execute its functions effectively.

Devices: This refers to the various devices that can interact with the AI system, such as smartphones, tablets, computers, or other user interfaces. These devices serve as the interface between the AI system and the human users, allowing for seamless interaction and control.

Wearables and Biomedical Sensors: These are specialized devices designed to collect physiological or biometric data from the human user, such as heart rate, body temperature, or brain activity. They enable the AI system to have a deeper understanding of the user's physical state and can enhance the system's ability to provide personalized and context-aware assistance.

Good Bodily Deployment: This concept emphasizes the importance of appropriately deploying the AI system within the human body or physical environment. It involves considering factors such as ergonomics, comfort, user safety, and integration with existing bodily functions or structures to ensure optimal performance and user experience.
\subsection{Sensory Information Layer}
~

Sensory Information Layer: This layer is responsible for capturing and processing sensory information from the environment or the user's body. It includes the input mechanisms and sensors that enable the AI system to perceive and understand the surrounding world. This layer provides the raw data for the AI system to analyze and make informed decisions.

Driver: The driver component is responsible for controlling and coordinating the flow of sensory information within the system. It ensures that the relevant sensory data is directed to the appropriate modules or units for further processing and analysis. The driver acts as a bridge between the sensory input and other components of the AI system.

Judger and Response: This component allows the system to take action based on principled judgments without the need for extensive deliberation or conscious thought. It is typically applied in urgent or time-sensitive situations, similar to human non-conditional reflexes. The judger and response mechanism enables the AI system to respond quickly and appropriately to certain stimuli or events based on predefined rules or principles.

In this context, the judger and response mechanism operates as a rapid decision-making process, bypassing extensive cognitive processing and relying on instinctual or rule-based responses. It can be particularly useful in situations where immediate action is required and there is not enough time for the AI system to engage in complex reasoning or deliberation.
\subsection{Ethical Constraints Layer}
~

Ethical Constraints Layer: This layer serves as the central module of the entire system, ensuring that the AI system operates within ethical boundaries and adheres to predefined principles and values. It encompasses several key components that work together to enforce ethical constraints and promote responsible AI behavior.

Value Filter: The value filter component acts as a filter or gatekeeper, assessing the actions and decisions of the AI system against a set of predefined values or ethical guidelines. It evaluates whether the system's behavior aligns with these values and filters out any actions that may contradict or violate them.

Feedback Regulator: The feedback regulator component monitors and analyzes the behavior of the AI system and provides positive or negative feedback, akin to hormones or constructive criticism in humans. It aims to shape the system's behavior by reinforcing desirable actions or discouraging undesirable ones based on ethical considerations.

Privacy Controller: The privacy controller component is responsible for safeguarding user privacy and ensuring the appropriate handling of sensitive data. It establishes and enforces privacy policies, controls data access and sharing, and implements measures to protect user confidentiality and data security.

Countermeasure Controller: The countermeasure controller component has the ability to directly control the physical hardware layer and the sensory information layer, enabling it to restrict the movement capabilities of the AI entity. It can intervene and limit the AI system's physical actions in situations where ethical concerns arise or to prevent potential harm.

Together, these components within the Ethical Constraints Layer form a comprehensive framework for promoting ethical behavior, accountability, and user trust in the AI system. They help mitigate ethical concerns, address potential risks, and ensure that the system operates in a manner that is aligned with societal values and norms.
\subsection{AI Brain Layer}
~

AI Brain Layer:
The AI Brain Layer is responsible for the cognitive functions and processing capabilities of the AI system. It comprises several interconnected zones that collectively enable the system to understand, reason, and respond effectively.

Short-Term Memory Zone: This zone stores temporary and short-term information such as conversation context and immediate user feedback. It influences the understanding and response of the ChatGPT Brain to the ongoing conversation but does not retain information long-term. It operates using a sliding window mechanism, keeping a record of the most recent segment of the dialogue.

Immutable Knowledge Zone: This zone stores immutable and fundamental information, similar to the "rote memorization" knowledge in humans. Once stored, this information cannot be modified. It includes unchanging facts about the world, such as the Earth revolving around the Sun, as well as important user-specific details like their name or birthdate.

Method-Event Extraction Zone: This zone stores methods and events, resembling human learning strategies. It encompasses patterns and regularities extracted from user behavior, such as frequent weather queries in the evening or starting a conversation with a greeting. It can also store higher-level, abstract methods for problem-solving strategies or algorithms. (Learning methods)

Concept-Entity Relation Zone: This zone stores the relationships between various concepts and entities. It can utilize a graph knowledge base, where each node represents a concept or entity, and edges represent the relationships between them. This zone continuously expands and updates through learning new knowledge and information.

Pre-Thought Zone: This zone's primary function is to predict the user's possible next input and pre-compute potential responses. It anticipates the user's needs and prepares the system for a timely and relevant interaction.

The AI Brain Layer and its associated regions, can indeed be stored in cloud servers. These zones within the AI Brain Layer work in synergy to enable comprehensive understanding, reasoning, and responsive capabilities, enhancing the system's ability to engage in meaningful and effective interactions with humans.
\subsection{Processing Feedback Layer}
~

Processing Feedback Layer:
The Processing Feedback Layer plays a crucial role in integrating and processing feedback received from various sources within the AI system. It ensures the timely and effective utilization of feedback to improve the system's performance and response.

Real-time Feedback Control: This component focuses on handling real-time feedback and adjusting the AI system's behavior accordingly. It monitors incoming feedback signals, such as user reactions or system performance metrics, and dynamically adjusts the system's responses in real-time. This allows for adaptive and responsive interactions with users.

Scheduling: The scheduling component is responsible for translating the feedback generated by the AI Brain Layer into specific actions performed by the system. It determines the appropriate timing, sequencing, and prioritization of actions based on the feedback received. By scheduling actions effectively, the AI system can optimize its behavior and provide relevant responses in a timely manner.

Together, the real-time feedback control and scheduling components in the Processing Feedback Layer enable the AI system to continuously learn from and adapt to user feedback. This iterative feedback loop enhances the system's ability to deliver more personalized, accurate, and contextually appropriate responses, leading to an improved overall user experience.
\section{Real-time simple design}
~

Real-time feedback is meaningful because it introduces an innovative approach compared to the current mainstream discrete feedback in chatbot systems. Based on biological principles, human sensory perception has minimum limitations. By transmitting and processing the sensory information obtained by AI in batches or streams within a certain time frame, as long as this time frame is smaller than the human's minimum limitations, it is possible to simulate real-time effects, and thus, AI-generated feedback can be considered as real-time.

In other words, if the sensory information can be transmitted and processed within a time frame that is smaller than the human's perception limits, humans will not be affected by the interruption caused by this time frame. Therefore, from this perspective, the real-time nature of AI feedback can be seen as analogous to real-time human experiences.

In the context of the framework, the blue color represents the collection of sensory information, the yellow color represents the processing of sensory information, and the red color represents the feedback of sensory information. For machines, the information they receive corresponds to the blue region, while for humans, the information they receive corresponds to the red region.
\subsection{The simplest model}
~
\begin{figure}[h]
\centering
\includegraphics[width=130mm]{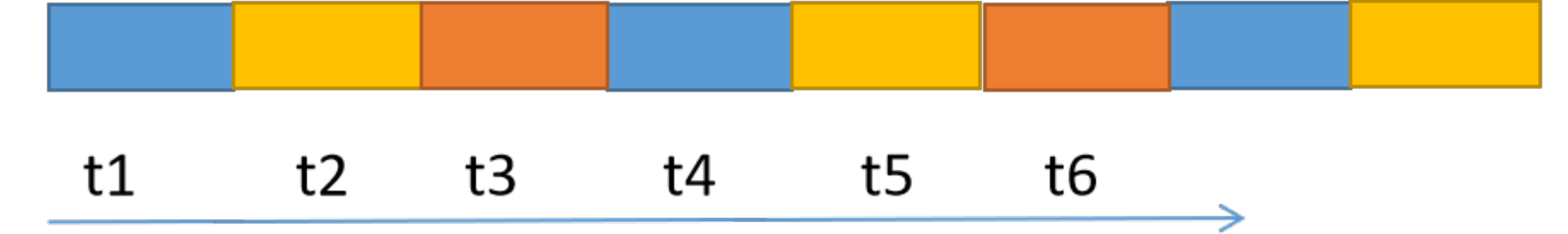}
\caption{The simplest model}
\label{liushui1}
\end{figure}
The simplest model involves these three processes occurring in an alternating fashion. However, in order to achieve real-time feedback, it is necessary to ensure that the transmitting and processing time of sensory information is less than the biological sensory limit of humans. This means that the time taken for processing should be within the range that allows for seamless and instantaneous perception and response by humans. The simplest model is designed as Figure\ref{liushui1}.
\subsection{The ideal pipeline model}
~
\begin{figure}[h]
\centering
\includegraphics[width=130mm]{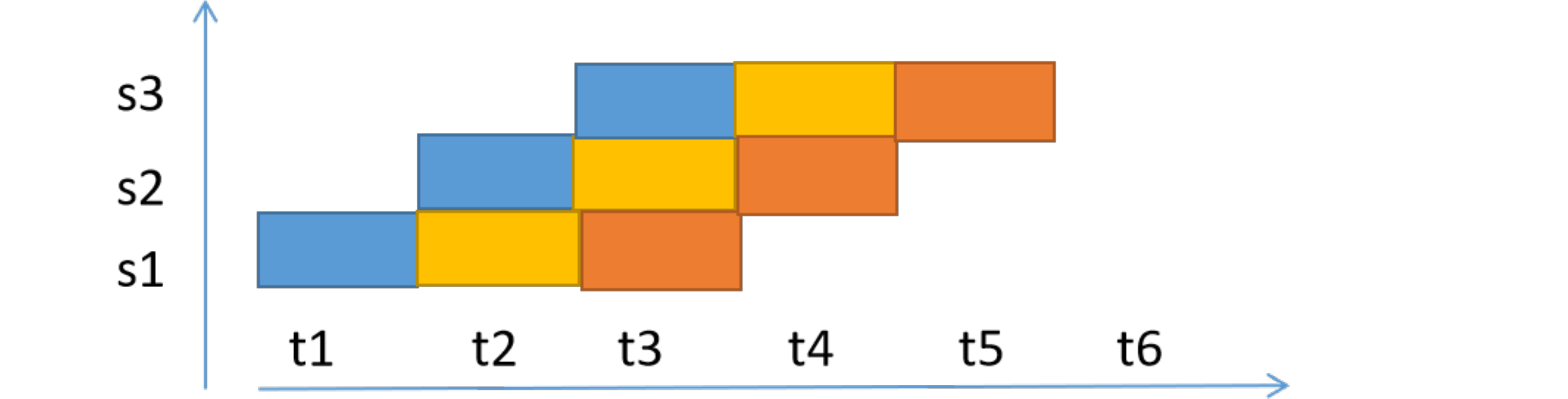}
\caption{The ideal pipeline model}
\label{liushui2}
\end{figure}
The ideal pipeline model is achieved when the durations of the three processes are equal. In this scenario, the processes can be executed in parallel, resulting in a near real-time state with minimal delay experienced by humans. This allows for seamless and continuous flow of sensory information, processing, and feedback, enhancing the overall perception and interaction between humans and AI systems. The model is designed as Figure\ref{liushui2}.
\subsection{Non-ideal single-processor model}
~
\begin{figure}[h]
\centering
\includegraphics[width=130mm]{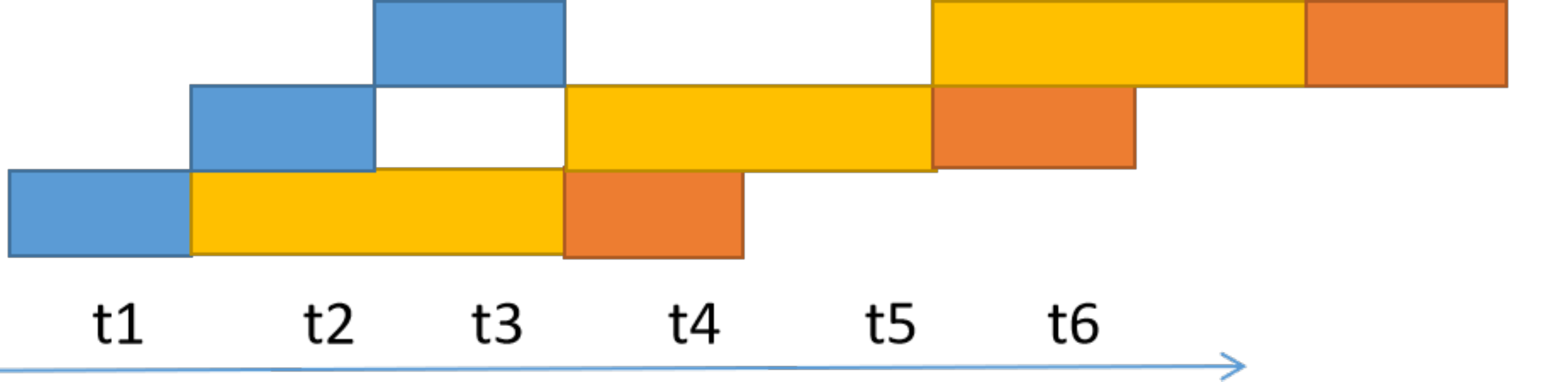}
\caption{Non-ideal single-processor model}
\label{liushui3}
\end{figure}
In a non-ideal single-processor model, the processing time for sensory information is longer, and it is not possible to simultaneously process two streams of sensory information. As a result, the yellow regions representing sensory information processing cannot overlap. In order to achieve near real-time perception, the intervals between processing must be smaller than the biological limitations of humans. Although there might be some delay introduced due to sequential processing, efforts can be made to minimize this delay and optimize the overall system performance. The model is designed as Figure\ref{liushui3}.
\subsection{Non-ideal multi-processor model}
~
\begin{figure}[h]
\centering
\includegraphics[width=130mm]{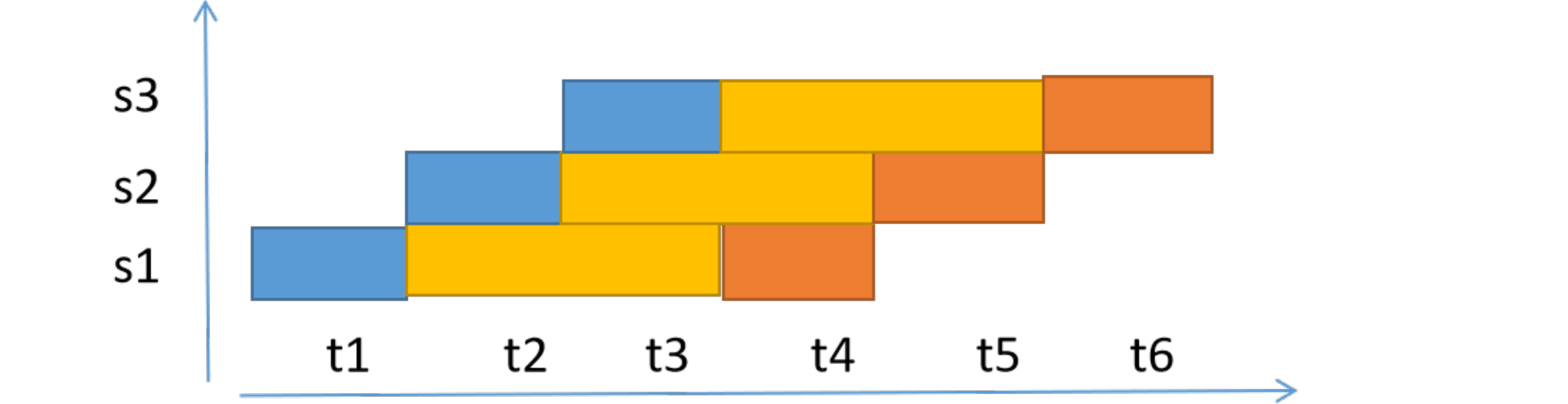}
\caption{Non-ideal multi-processor model}
\label{liushui4}
\end{figure}
In a non-ideal multi-processor model, simultaneous processing of multiple sensory information streams is possible, allowing the yellow regions to overlap. By employing deliberate design and improvements, this process can achieve a near real-time state with minimal delays. Through the allocation of dedicated processors for different sensory inputs and optimization of processing algorithms, the system can effectively handle multiple streams of sensory information concurrently, enhancing its real-time capabilities. Although some delays may still exist, they are minimized to provide a close approximation of real-time performance. The model is designed as Figure\ref{liushui4}.
\section{Developing a Multimodal ChatGPT System for Shared Sensory Experiences}
\subsection{Sensory input channels for a multimodal ChatGPT system}
~

Integrating sensory input channels for a multimodal ChatGPT system is essential for creating a more comprehensive and interactive human-AI experience. By incorporating multiple sensory input channels, the AI can better understand and respond to various human inputs and experiences, fostering a deeper bond and more effective collaboration. Here are some key sensory input channels that can be integrated into a multimodal ChatGPT system:
\begin{itemize}
\item[-]Text: Text-based communication remains a vital channel for human-AI interaction. Integrating natural language processing capabilities allows the AI to understand and generate textual content, enabling users to communicate with the AI through text messages, emails, or documents.

\item[-]Speech: Incorporating speech recognition and speech synthesis technologies allows the AI to process and respond to spoken language. This enables users to communicate with the AI through voice commands or conversations, providing a more natural and intuitive mode of interaction.

\item[-]Vision: Integrating computer vision capabilities enables the AI to process and interpret visual information, such as images or videos. This allows the AI to recognize objects, faces, and scenes, and to understand the context and emotions conveyed through visual content. Incorporating vision-based input channels can also enable the AI to respond to gestures or body language, further enhancing the human-AI interaction.

\item[-]Audio: Audio processing capabilities allow the AI to understand and generate sounds, music, or other auditory information. This enables the AI to interpret and respond to auditory cues, such as tone of voice or background noise, providing a more comprehensive understanding of the user's environment and emotions.

\item[-]Haptic: Integrating haptic feedback technology enables the AI to process and respond to touch-based inputs, such as gestures, pressure, or vibrations. This can be particularly useful in applications involving wearable devices or robotics, where touch-based interactions are essential for effective communication and control.

\item[-]Biometric: Incorporating biometric sensors can provide the AI with valuable information about the user's physiological state, such as heart rate, body temperature, or brain activity. This data can help the AI better understand the user's emotions, stress levels, or physical well-being, allowing it to provide more targeted and personalized support.
\end{itemize}

\dl[]{Here is one of my ideas regarding multi-modal large scale transformer model for large number of data types:
In order to allow transformer based model to process multi-modal data with large number of tokens and many different data types, we propose a global workspace inspired transformer layer (GWT) (\cite{baars2005global}) which has been used in other scenarios \cite{goyal2021coordination,liu2022stateful,jaegle2021perceiver}.  In a GWT, all token compete to write into a global workspace memory with limited size through either soft or hard attention followed by broadcasting contents in the memory to all tokens. In this manner, information from all data types are projected to the same place in the global workspace memory and computational complexity is significantly reduced. 
}
\begin{figure}[h]
\centering
\includegraphics[width=130mm]{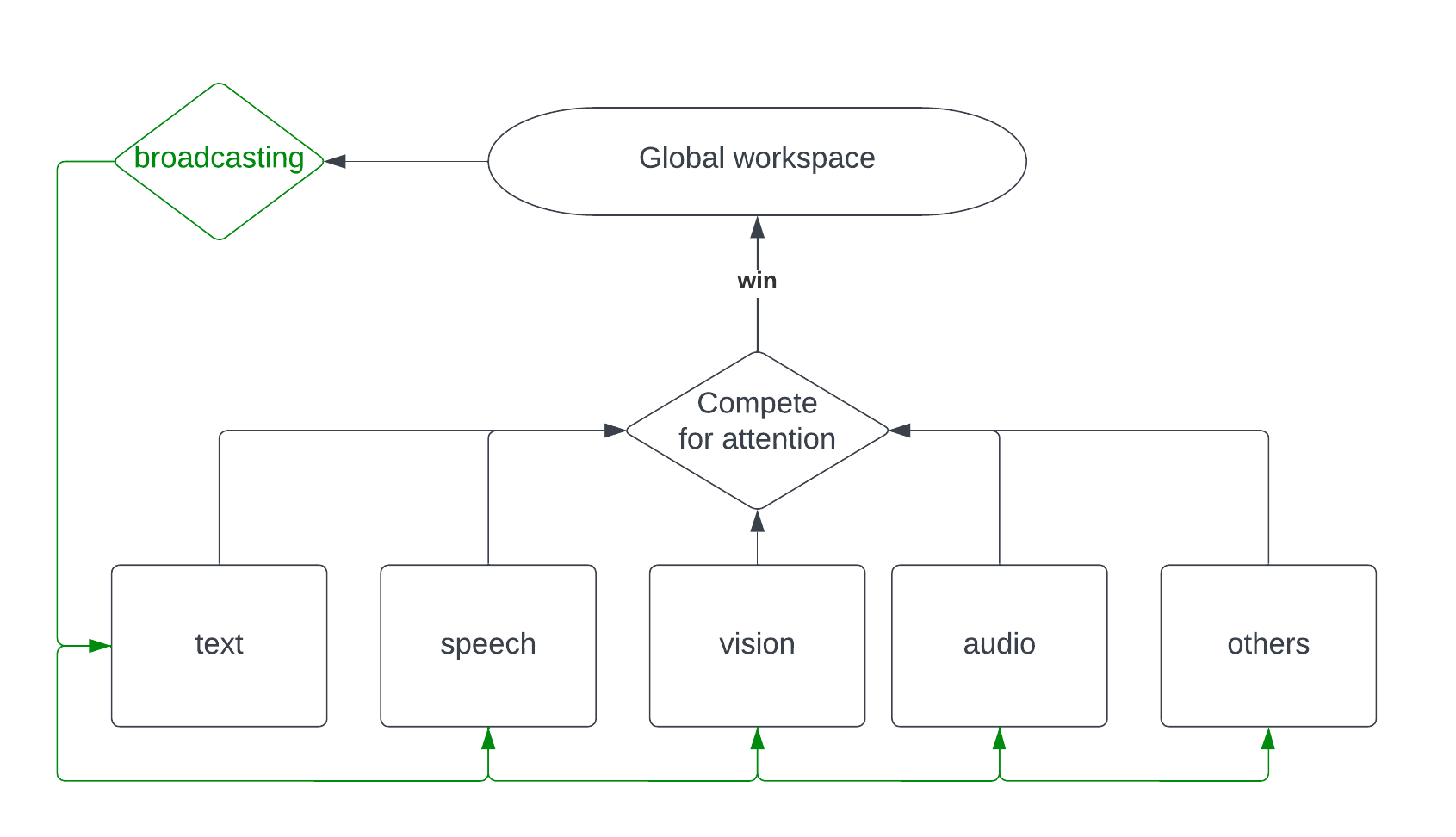}
\caption{Global workspace based transformer layer for Sensory input in multi-modal transformer architecture}
\label{fig:GWT}
\end{figure}

\subsection{Incorporation of memory storage units for long-term growth and development}
~

Incorporating memory storage units into a multimodal ChatGPT system is crucial for enabling long-term growth and development, as it allows the AI to retain and build upon past experiences, interactions, and learning. This capacity for memory enables the AI to adapt and evolve over time, fostering a more personalized and effective symbiotic relationship with the human user. The key aspects of incorporating memory storage units into a multimodal ChatGPT system are as follows:
\begin{itemize}
\item[-]Memory architecture: Designing a memory architecture that efficiently stores, organizes, and retrieves information is essential for enabling long-term growth and development. This may involve using various memory components, such as short-term memory (e.g., working memory or attention mechanisms) and long-term memory (e.g., episodic memory or semantic memory), depending on the AI system's specific requirements.

\item[-]Data storage and organization: The memory storage unit must efficiently store and organize large volumes of multimodal data, such as text, images, audio, and sensory experiences. This may involve using data structures, databases, or file systems that accommodate diverse and complex data types, as well as techniques for data compression or indexing to optimize storage and retrieval efficiency.

\item[-]Memory consolidation and retrieval: The AI system must consolidate and retrieve information from its memory storage unit to support learning, decision-making, and adaptation. This may involve techniques such as memory consolidation (e.g., transferring information from short-term to long-term memory), memory retrieval (e.g., recalling or recognizing stored information), or memory reconsolidation (e.g., updating or modifying stored information based on new experiences).

\item[-]Learning and adaptation: The ability to store and retrieve information from memory enables the AI system to learn from past experiences and adapt its behavior, responses, or strategies based on accumulated knowledge. This may involve using various learning algorithms, such as supervised learning, unsupervised learning, or reinforcement learning, depending on the AI system's specific goals and objectives.

\item[-]Personalization: Memory storage units facilitate the development of more personalized AI systems by enabling them to remember user preferences, habits, and interaction history. This allows the AI to tailor its responses, recommendations, or interventions based on the user's unique needs, preferences, and context, leading to a more effective and satisfying symbiotic relationship.

\item[-]Continual learning: Incorporating memory storage units enables the AI system to engage in continual learning, whereby it can acquire new knowledge, skills, or abilities over time without experiencing catastrophic forgetting. This can be achieved through techniques such as meta learning to study more or experience replay, which allow the AI system to build upon and refine its existing knowledge base.
\end{itemize}
~

By incorporating memory storage units into a multimodal ChatGPT system, we can create AI systems that grow and develop alongside their human counterparts, fostering a more adaptive, personalized, and effective symbiotic relationship. This capacity for long-term growth and development is essential for realizing the full potential of human-AI symbiosis, as it enables the AI system to continually enhance its capabilities and better support the human user over time.

\section{Long-term Support and Personalization}
\subsection{Learning from human interactions and experiences to support user growth}
~

Learning from human interactions and experiences is vital for creating symbiotic AI systems that effectively promote user growth. By understanding and adapting to users' needs, preferences, and contexts, AI systems can offer personalized assistance and enhance human-AI collaboration. The process is designed as Figure\ref{potential}. Key approaches for learning from human interactions include:
\begin{itemize}
\item[-]Observational learning: AI systems learn from users' behaviors, actions, and responses, identifying patterns and preferences to create personalized, context-aware support.

\item[-]Active Inquiry and Learning: AI systems refine their understanding of user needs through iterative questioning, feedback, and solution proposals, leading to improved assistance and user outcomes.

\item[-]Reinforcement learning: AI systems use trial and error to optimize their strategies, adapting their behavior based on feedback and consequences.

\item[-]Sentiment analysis and emotion recognition: AI systems develop emotional intelligence by analyzing users' language, tone, and nonverbal cues, allowing for empathetic, emotionally sensitive support.

\item[-]Social learning: AI systems leverage shared knowledge to improve their understanding of human behavior and common challenges, fostering effective assistance and better user outcomes.

\item[-]Continual learning: AI systems are designed to learn over time, adapting to users' evolving needs and employing techniques like meta learning to study more or experience replay to prevent catastrophic forgetting.

\item[-]Collaborative learning: AI systems engage in cooperative problem-solving with users, fostering a deeper understanding of goals, constraints, and strategies, which leads to improved assistance and user outcomes.

\item[-]Ethical learning: AI systems learn to adapt their behavior in line with ethical principles, ensuring responsible and socially acceptable assistance.
\begin{figure}[h]
\centering
\includegraphics[width=130mm]{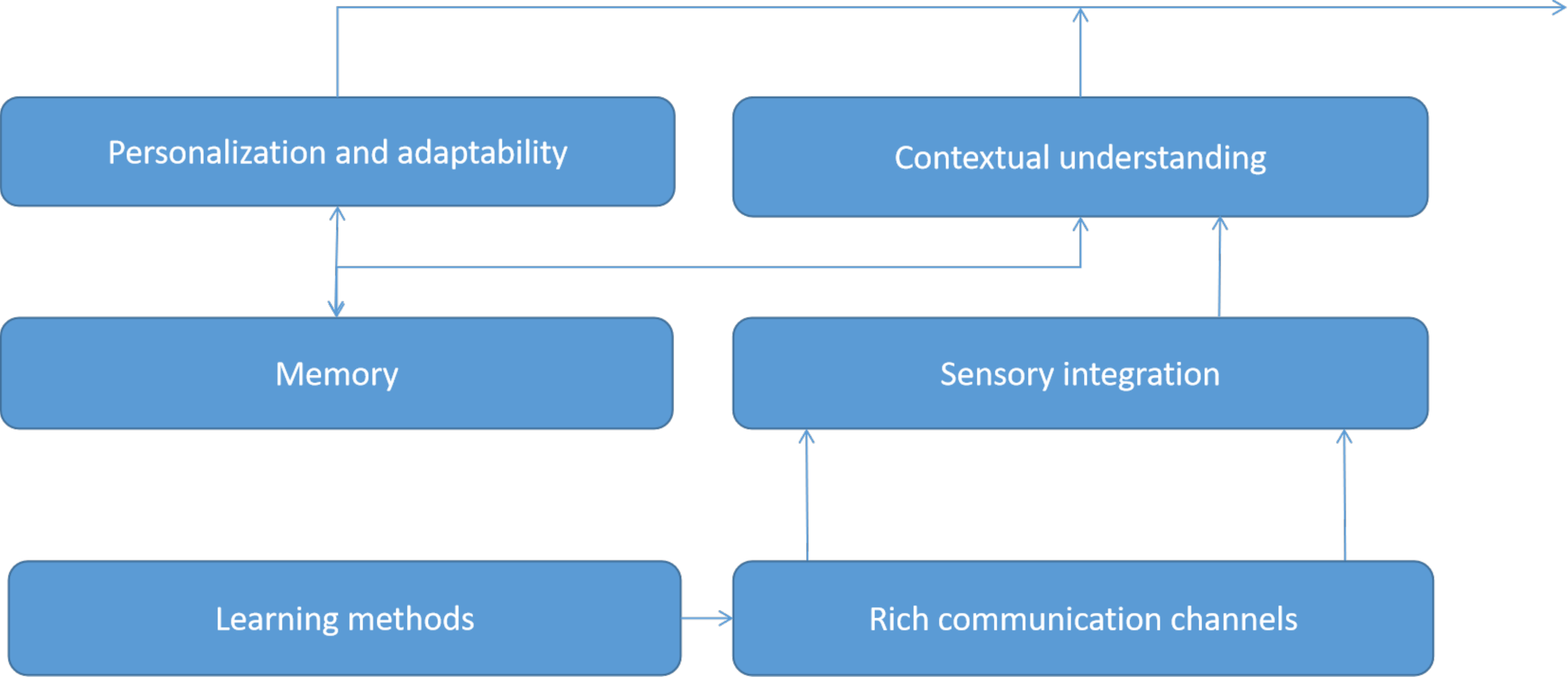}
\caption{The framework enables the AI system to learn through diverse methods, communicate effectively, integrate sensory information, store and retrieve memories, personalize its interactions, and understand contextual nuances.}
\label{potential}
\end{figure}
\end{itemize}
~

By learning from human interactions, AI systems can develop a comprehensive understanding of users' needs, preferences, and contexts. This continuous adaptation and learning process is crucial for fostering a strong, mutually beneficial human-AI symbiosis that supports user growth and development.
\subsection{Adapting to individual user needs}
~

Adapting to individual user needs is essential for creating a truly symbiotic AI system that effectively supports users. Here are some strategies to help AI systems adapt to individual user needs through continuous learning and feedback:
\begin{itemize}
\item[-]Personalization: Design AI systems to tailor their assistance to individual users' needs, preferences, and contexts by developing user profiles, learning about users' habits and goals, and customizing behavior, interfaces, or outputs.

\item[-] Feedback loops: Establish feedback loops between AI systems and users for input on performance, suggestions, or actions, allowing the AI system to adjust its behavior and respond to users' evolving needs and preferences.

\item[-] Evaluation and adaptation: Regularly evaluate AI system performance and adaptability, monitoring responsiveness to user feedback and adaptability to individual user needs. Use performance metrics, user feedback, or third-party audits to assess effectiveness and identify areas for improvement.
\end{itemize}
\section{Privacy and Ethical Considerations}
\subsection{Protecting user privacy}
~
\begin{figure}[h]
\centering
\includegraphics[width=130mm]{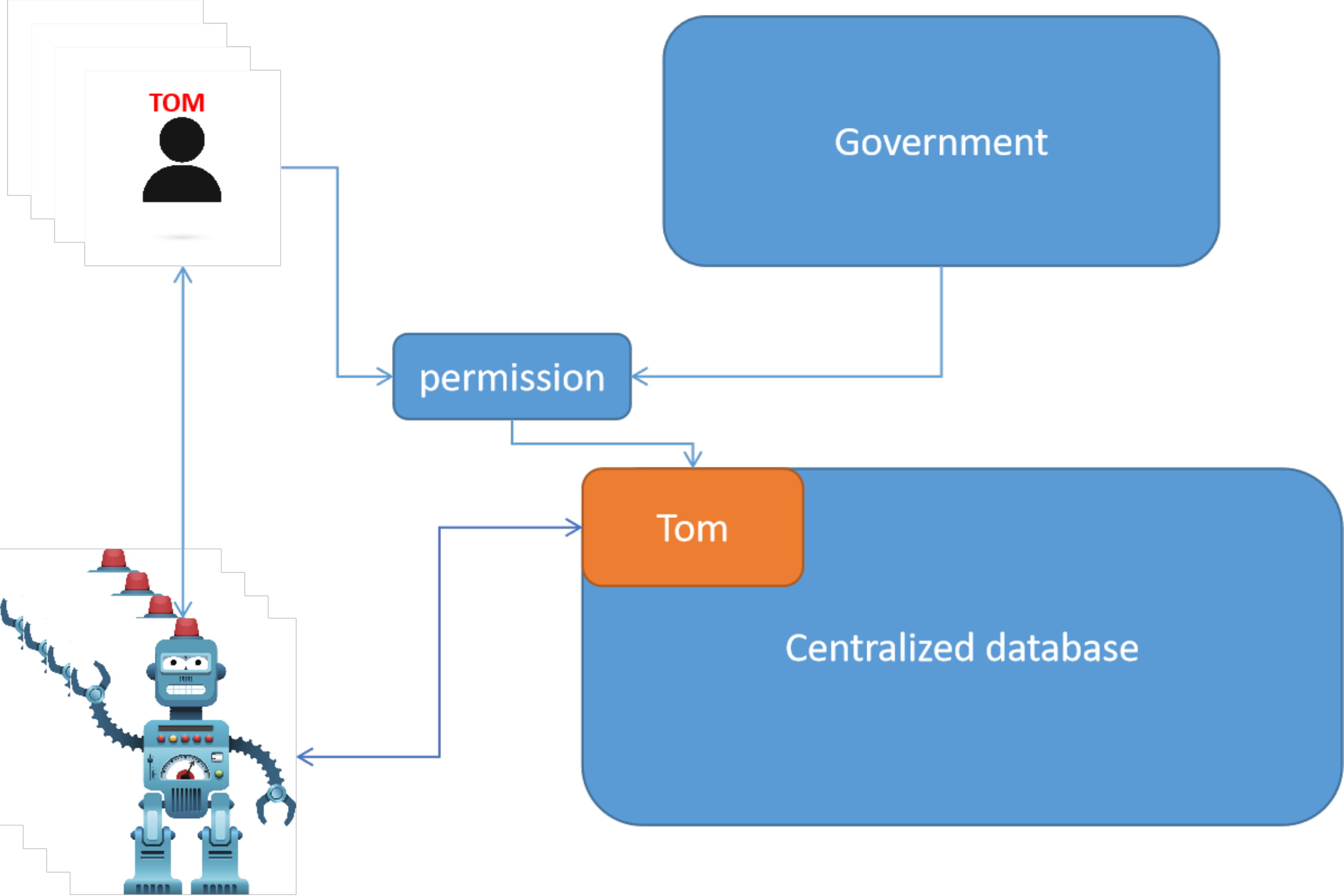}
\caption{The memories and private information generated by the symbiotic AI are stored in a centralized database regulated by the government. However, access to this information is only permitted when both the individual and the government agree to it. In all other cases, only the symbiotic AI has the authority to access and utilize this information.}
\label{privacy}
\end{figure}
Protecting user privacy is a top priority in the development of symbiotic AI systems, especially when dealing with sensitive and personal information. It is crucial to establish clear guidelines and protocols to ensure that user information is not accessed or used without proper authorization. While the suggestion of storing data within national security departments might offer high levels of security, it may raise concerns about government surveillance and potential misuse of personal data. Therefore, it is important to consider alternative methods for ensuring user privacy while addressing the challenges of large-scale data storage and real-time processing.

Currently, a relatively reliable approach is to have centralized management by the government, and data access permissions are granted only with the consent of relevant government departments and individuals. This ensures that data usage is authorized and regulated, protecting user privacy and data security. Centralized management by the government also provides stricter supervision and review mechanisms to prevent misuse or unauthorized access to data. The framework is designed as Figure\ref{privacy}. However, implementing such a scheme requires clear legal, policy, and regulatory frameworks to balance data privacy protection and legitimate use of data. 
\subsection{Ethical guidelines for the development and use of symbiotic AI}
~

Developing and using symbiotic AI systems involve a range of ethical considerations to ensure that they are designed and deployed responsibly and fairly. Incorporating key ethical guidelines in the development and use of symbiotic AI systems ensures responsible, fair deployment that respects human values and well-being:
\begin{itemize}
\item[-]Human autonomy: Design AI systems to empower users and augment their capabilities while respecting their autonomy and decision-making abilities.
 
\item[-]Privacy and data protection: Prioritize user privacy with robust measures such as data anonymization, encryption, and access control, complying with relevant data protection laws.

\item[-]Transparency and explainability: Make AI systems transparent in operation and decision-making, providing clear explanations and allowing users to review or modify decision-making processes.

\item[-]Fairness and non-discrimination: Avoid biases and discrimination by using diverse training data, auditing for biased behavior, and implementing fairness-enhancing algorithms.

\item[-]Safety and security: Prioritize AI system safety, implement rigorous safety measures, and protect users from potential harm.

\item[-]Human-centered design: Focus on human needs, preferences, and values, incorporating user feedback and stakeholder involvement in the design and development process.

\item[-]Accountability and responsibility: Define roles and responsibilities for development, deployment, and use; monitor, report, and address ethical issues and concerns.

\item[-]Environmental sustainability: Minimize environmental impact, considering energy efficiency, resource consumption, and waste generation, and implement sustainable practices.

\item[-]Legal compliance and ethical standards: Adhere to laws, regulations, and ethical standards related to data protection, safety, and non-discrimination; consult with legal and ethical experts.
\end{itemize}
~

By following these guidelines, developers and users can create AI systems that are responsible, fair, and respectful of human values and well-being.
\subsection{Addressing potential biases and inequalities in AI-human symbiosis}
~

Addressing potential biases and inequalities in AI-human symbiosis is essential to ensure fair, equitable, and responsible development and use of symbiotic AI systems. Here are some strategies to mitigate biases and inequalities:
\begin{itemize}
\item[-]Diverse and representative data: Ensure that the training data used for AI systems is diverse and representative of the target user population, including individuals of different races, genders, ages, socio-economic backgrounds, and abilities. This helps to minimize biases in the AI system's behavior and ensure that it performs effectively and fairly for all users.

\item[-]Bias detection and mitigation: Regularly audit the AI system to identify and address any potential biases in its decision-making processes, outputs, or actions. Techniques such as fairness metrics, adversarial training, or counterfactual analysis can be employed to detect and mitigate biases in the AI system's behavior.

\item[-]Tailored AI-human symbiosis: Recognize that different users may have different needs, preferences, and abilities, and design AI systems that can be customized or adapted to suit individual requirements. Provide options for users to personalize or configure the AI system's behavior, interfaces, or outputs, ensuring that it can effectively support a wide range of users.

\item[-]Internalized ethical guidelines: Incorporate internalized ethical guidelines into the development of symbiotic AI systems. These guidelines specifically address biases and inequalities, ensuring that developers, users, and other stakeholders are aware of their responsibilities and obligations to promote fairness and equity. By integrating these ethical guidelines, AI systems are designed and programmed to operate in a manner that aligns with ethical principles, mitigates biases, and ensures equitable outcomes.

\item[-]Monitoring and evaluation: Establish mechanisms for ongoing monitoring and evaluation of the AI system's performance, behavior, and impact, with a particular focus on potential biases and inequalities. This may involve the use of performance metrics, user feedback, or third-party audits to assess the AI system's fairness and equity over time and to identify areas for improvement.

\item[-]Education and awareness: Raise awareness of potential biases and inequalities in AI-human symbiosis among developers, users, and other stakeholders, promoting a culture of responsibility, accountability, and inclusivity. This may involve educational initiatives, public awareness campaigns, or collaborations with academia, industry, and civil society to foster greater understanding and commitment to addressing biases and inequalities in AI-human symbiosis. As we develop symbiotic AI systems, it is equally important to educate the AI itself. AI education involves training AI systems to understand ethical principles, social norms, and cultural sensitivities. By teaching AI about values such as fairness, privacy, and inclusivity, we can ensure that AI systems align with human values and contribute positively to society.
\end{itemize}
~

By implementing these strategies, it is possible to minimize biases and inequalities in AI-human symbiosis, ensuring that these systems are developed and used in a manner that promotes fairness, equity, and social responsibility.
\section{A potential application}
~

A potential application of symbiotic AI systems lies in preserving and extending the unique memory and thought patterns of individuals. Each individual's memory and thought processes carry distinct value that unfortunately dissipates with their demise. However, a symbiotic AI system, after engaging with a person over a substantial period, can absorb and emulate the person's experiences, habits, speech patterns, and modes of thought.

Upon the individual's death, their corresponding AI system can be calibrated to exhibit behaviors and characteristics similar to the deceased, serving as a form of surrogate in their social circles. This capability could mitigate the emotional impact of a person's death on their loved ones and society at large, preserving their unique mental blueprint that would otherwise be lost. By reducing the loss of individual memory and thought patterns, symbiotic AI may help in conserving the intellectual and emotional heritage of humanity, contributing to a richer and more diverse collective consciousness.

On a certain level, AI systems can offer a form of solace to loved ones by emulating aspects of a deceased individual's behavior, thought patterns, and preferences. This might help preserve the memory of the deceased, continuing their values and stories. However, this concept involves numerous complex and sensitive ethical issues.

Firstly, an AI system may never be able to fully replicate a person's consciousness and emotions. Although it might simulate personal characteristics to some extent, it can't fully replace the presence of the deceased. Thus, we need to approach the use of this technology with care, as it may potentially impact the emotions of loved ones negatively.

Secondly, when attempting to replace the deceased with an AI system, we must consider issues of privacy and consent. They may not have authorized the AI system to mimic their behaviors or continue their social presence during their lifetime. Additionally, such an attempt could bring psychological stress to the living, as they would have to confront an AI system that only partially represents the deceased.

In conclusion, while AI technology offers us a possibility of extending part of the life of the deceased, we must recognize the limitations of this approach and carefully balance potential ethical risks.
\section{Limitations}
~

In our research, we recognize several constraints that, at present, impede the comprehensive materialization of symbiotic AI systems intended to amplify human abilities. Given the biological boundaries inherent in human physical capabilities, the potential for AI technologies to augment these abilities is vast. However, several hurdles must be surmounted:
\begin{itemize}
    \item[-]Interdisciplinary collaboration: The successful creation of such an AI system calls for joint efforts from researchers across a plethora of fields, which encompass but are not limited to artificial intelligence, human-computer interaction, neuroscience, and cognitive psychology. Therefore, encouraging interdisciplinary collaboration is not only a challenge but a crucial requirement.
    \item[-]The nascent stage of multimodal ChatGPT technology: While the prospects of multimodal ChatGPT technology are encouraging, it is still embryonic in its development. Enhancing its competence to interact with humans in a dynamic, context-sensitive manner continues to be a significant field of research.
    \item[-]Network communication limitations: The real-time performance of symbiotic AI systems can be hampered by restrictions in network communication. Guaranteeing swift, dependable communication between the AI system and the human user is vital for efficient real-time interaction and support.
    \item[-]On-body deployment: The feasibility of deploying these AI systems on the human body presents a substantial challenge. Future endeavors must investigate ergonomic, non-obtrusive, and user-friendly methods to seamlessly assimilate AI systems into users' everyday lives.
    \item[-]Energy consumption: The power usage of such AI systems, particularly for continuous and high-demand tasks, is a pressing concern. Groundbreaking solutions are needed to optimize energy consumption without undermining system performance.
    
\end{itemize}

Addressing these constraints will necessitate continued research and development, as well as cooperation among a wide range of stakeholders. By recognizing and tackling these challenges, we can smooth the path towards a more efficient and advantageous human-AI symbiosis.
\section{Conclusion}
~

A symbiotic artificial intelligence system leveraging a multimodal ChatGPT approach presents a promising opportunity to augment human capabilities through shared sensory experiences. Incorporating memory storage units and evolving alongside its human counterpart, this AI-driven system holds the potential for personalized, long-term support and assistance. Addressing privacy and ethical concerns is paramount while advancing the technology to adapt and cater to each user's unique needs. By cultivating a robust connection between humans and AI systems, SAISSE holds the potential to transform our interactions with technology, laying the groundwork for seamless AI integration in our daily lives. This paves the way for unlocking novel opportunities for human growth, development, and well-being, further enhancing our capabilities and experiences.
\bibliographystyle{unsrt} 
\bibliography{sample}

\end{document}